\documentclass[paper,twocolumn,twoside]{geophysics}

\usepackage{graphicx}
\usepackage{amsfonts}
\usepackage{amsmath}
\usepackage{epstopdf}
\usepackage{bm}
\usepackage{caption}
\usepackage{alphalph}
\usepackage[table]{xcolor}
\usepackage{tikz}
\usepackage{pgfplots}
\usepackage{subfig}
\usepackage{url}
\usepackage{booktabs}
\usepackage{multirow}
\usepackage{diagbox}
\usepackage{siunitx}
\usepackage{hyperref}
\usepackage{alphalph}
\usepackage{soul} 

\definecolor{upperns}{RGB}{69,117,180}
\definecolor{middlens}{RGB}{145,191,219}
\definecolor{lowerns}{RGB}{224,243,248}
\definecolor{rijnchalk}{RGB}{254,224,144}
\definecolor{scruff}{RGB}{252,141,89}
\definecolor{zechstein}{RGB}{215,48,39}

\newcommand{\hlc}[2][yellow]{ {\sethlcolor{#1} \hl{#2}} }

\begin{document}

\title{A Machine Learning Benchmark for Facies Classification}

\renewcommand{\thefootnote}{\fnsymbol{footnote}} 

\address{
\noindent \footnotemark[1] Corresponding author, email: \texttt{alaudah@gatech.edu} \\
\footnotemark[3] Center for Energy and Geo Processing (CeGP) at 
Georgia Institute of Technology and King Fahd University of Petroleum \& Minerals (KFUPM).  
\footnotemark[2] Faculty of Earth Sciences, 
University of Silesia, 
Katowice, Poland.}
\author{Yazeed Alaudah\footnotemark[1]\footnotemark[3], Patrycja Micha\l{}owicz\footnotemark[2], Motaz Alfarraj\footnotemark[3], and Ghassan AlRegib\footnotemark[3]}

\lefthead{Alaudah \textit{et al.}}
 
\begin{abstract}

The recent interest in using deep learning for seismic interpretation tasks, such as facies classification, has been facing a significant obstacle, namely the absence of large publicly available annotated datasets for training and testing models. As a result, researchers have often resorted to annotating their own training and testing data. However, different researchers may annotate different classes, or use different train and test splits. In addition, it is common for papers that apply machine learning for facies classification to not contain quantitative results, and rather rely solely on visual inspection of the results. All of these practices have lead to subjective results and have greatly hindered the ability to compare different machine learning models against each other and understand the advantages and disadvantages of each approach. 

To address these issues, we open-source a fully-annotated 3D geological model of the Netherlands F3 Block. This model is based on the study of the  3D seismic data in addition to 26 well logs, and is grounded on the careful study of the geology of the region. Furthermore, we propose two baseline models for facies classification based on a deconvolution network architecture and make their codes publicly available. Finally, we propose a scheme for evaluating different models on this dataset, and we share the results of our baseline models. In addition to making the dataset and the code publicly available, this work helps advance research in this area by creating an objective benchmark for comparing the results of different machine learning approaches for facies classification.

\end{abstract}

\section{Introduction}
In recent years, there has been great interest in using fully-supervised deep learning models for seismic interpretation tasks such as facies classification \cite[]{shi_wu_fomel_seg2018, dramsch_seg2018, waldeland2017salt,araya2017automated,huang2017scalable, tao_zaho_seg2018, haibin_deconv_seg2018, Charles2017MalenoV}. Typically, deep learning models --- such as convolutional neural networks (CNNs) --- have millions of free parameters, and therefore require a large amount of annotated training data. Unfortunately, and unlike other areas of research such as computer vision, there is a lack of large publicly-available annotated datasets for seismic interpretation that can be used to train and benchmark machine learning models. To address this problem, some researchers resort to annotating their own training and testing datasets. For example, in the Netherlands F3 block, \cite{tao_zaho_seg2018} annotated 40 inlines, \cite{haibin_deconv_seg2018} annotated 12 inlines, while \cite{Charles2017MalenoV} only annotated a single inline for their model. The limited number of annotated sections is understandable given that the annotation process is time-consuming, requires subject matter expertise, and can be quite subjective. Nevertheless, such limited annotations undermine the mass potential machine learning could have when deployed in such a field.  

Alternatively, there has been some research in attempting to avoid annotating large amounts of data by using weakly-supervised learning approaches. \cite{icip2016} trained a facies classification model using seismic images with image-level labels only. Later,  \cite{alaudah2018geophysics} proposed a method for generating large amounts of training data using similarity-based retrieval and a weakly-supervised label mapping algorithm. As few as one or two exemplar images per class were enough to automatically generate a large amount of training data. These automatically-generated training data were then used to train a weakly-supervised deconvolution network \cite[]{yazeed_seg2018_label} for facies classification. Other researchers avoid supervision all-together by using traditional unsupervised machine learning techniques such as principal component analysis or self-organizing maps. There is a very rich literature on traditional supervised and unsupervised methods for facies classification, e.g., \cite[]{coleou2003unsupervised,de2006unsupervised,DUBOIS2007599}. \cite{zhao_review_paper} provides a review of some of the most commonly used techniques. More recently, unsupervised techniques based on deep learning models such as deep convolutional autoencoders have been explored \cite[]{qian_et_al_geophysics_2018, mohit2018, veillard2018fast}. 

Whether researchers annotate their own training data or use other techniques, there still remains a lack of large publicly-available annotated datasets for seismic interpretation that can be used for training different models and comparing the performance of different approaches. Furthermore, it is common for papers that apply machine learning for facies classification, or other seismic interpretation tasks, to not contain quantitative results, but rather rely solely on subjective visual inspection of the results. All of this leads to highly subjective results and greatly hinders the ability of researchers to compare different approaches against each other and understand the advantages and disadvantages of each approach. 

To address these issues, and to help make machine learning research in seismic interpretation more reproducible, we open-source a fully-annotated 3D geological model of the Netherlands F3 Block \cite[]{F3_data}. This model is grounded in the geology of the region and based on the study of both the 3D seismic data and 26 well logs located within the F3 block or its vicinity. The data also includes fault planes that we have extracted from the F3 block. While we do not use the fault data in this work, we make the data publicly available for those who are interested in exploring fault detection within our model. Furthermore, we also present two baseline models for facies classification based on a deconvolution network architecture. The first baseline is a patch-based model that is trained using a large number of small patches extracted from all the inlines and crosslines of the training set. The second baseline is a section-based model that was trained directly on entire inlines and crosslines of the training set. In addition, we have open-sourced all the codes that were used to train and test our baseline models using the \textsc{PyTorch} deep learning library\footnote{Both the code and the dataset can be viewed at: 
\href{https://github.com/olivesgatech/facies_classification_benchmark}{\texttt{www.github.com/olivesgatech/facies\_classification\_benchmark}}}. Finally, we propose a common procedure for evaluating different models on this dataset, and we share the results for our baseline models. The next section provides a quick overview of the geology of the Netherlands F3 block and introduces our geological model.

\section{A 3D Geological Model of the Netherlands F3 Block}
\begin{figure*}[!ht]

\begin{center}
\includegraphics[width=0.8\textwidth]{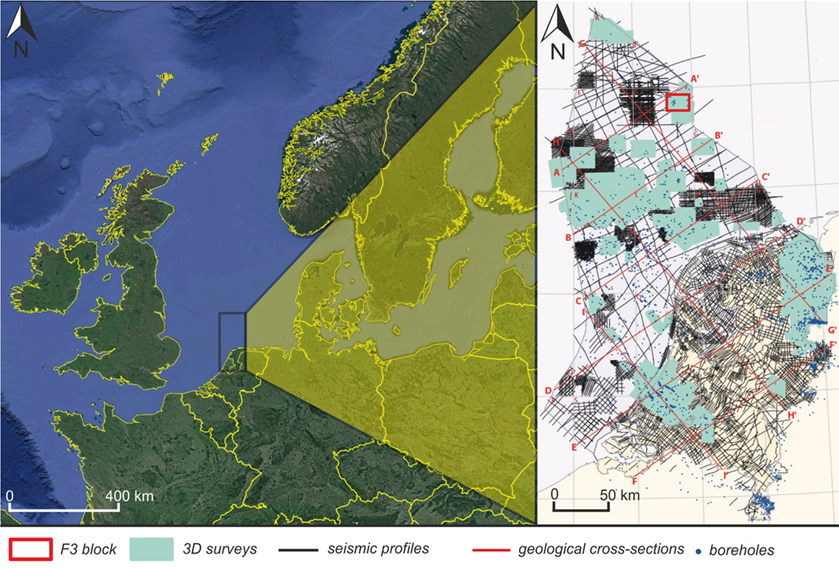}
\end{center}

\caption{The location of the F3 block. Adapted from \cite{duin2006subsurface}.}
\label{fig:f3}
\end{figure*}

The North Sea is rich in hydrocarbon deposits, which is why this area is very well studied in the literature \cite[]{doornenbal2014kilka}. The North Sea continental shelf, located off the shores of the Netherlands, is divided into geographical zones described by different letters of the alphabet; within these zones are smaller areas marked with numbers. One of these areas is a rectangle of dimensions 16 km x 24 km known as the F3 block, see Figure \ref{fig:f3}. In 1987, the F3 block 3D seismic survey was conducted to identify the geological structures of this area and to search for hydrocarbon reservoirs. In addition, many boreholes were drilled within the F3 block throughout the years. The F3 block became one of the most widely known and studied seismic surveys after dGB Earth Sciences made the data obtained from the survey publicly available. 

The aim of this section is to briefly describe the geology of the survey area and introduce the 3D geological model that we have developed and how it was obtained. 

\subsection{The geology of the F3 block}
Within the continental shelf of the North Sea, ten groups of lithostratigraphic units have been identified in the literature \cite[]{van19931997, mijnlieff_2002, scheck2005crustal, duin2006subsurface}. These groups and their main lithostratigraphic features are listed below from newest to oldest:  
\begin{enumerate}
    \item Upper North Sea group: claystones and sandstones from Miocene to Quaternary.

    \item Lower and Middle North Sea groups: sands, sandstones, and claystones from Paleocene to Miocene.

    \item Chalk group: carbonates of Upper Cretaceous and Paleocene.

    \item Rijnland group: clay formations with sandstones of Upper Cretaceous.

    \item Schieland, Scruff and Niedersachsen groups: claystones of Upper Jurassic and Lower Cretaceous.

    \item Altena group: claystones and carbonates of Lower and Middle Jurassic.

    \item Lower and Upper Germanic Trias groups: sandstones and claystones of Triassic.

    \item Zechstein group: evaporites and carbonates of Zechstein.

    \item Upper and Lower Rotliegend groups: siliceous rocks and basalts of the Lower Zechstein.

    \item Limburg group: Upper carboniferous siliceous rock, which are the bedrock for hydrocarbons.
\end{enumerate}

\begin{figure*}[t]

\begin{center}
\includegraphics[width=\textwidth]{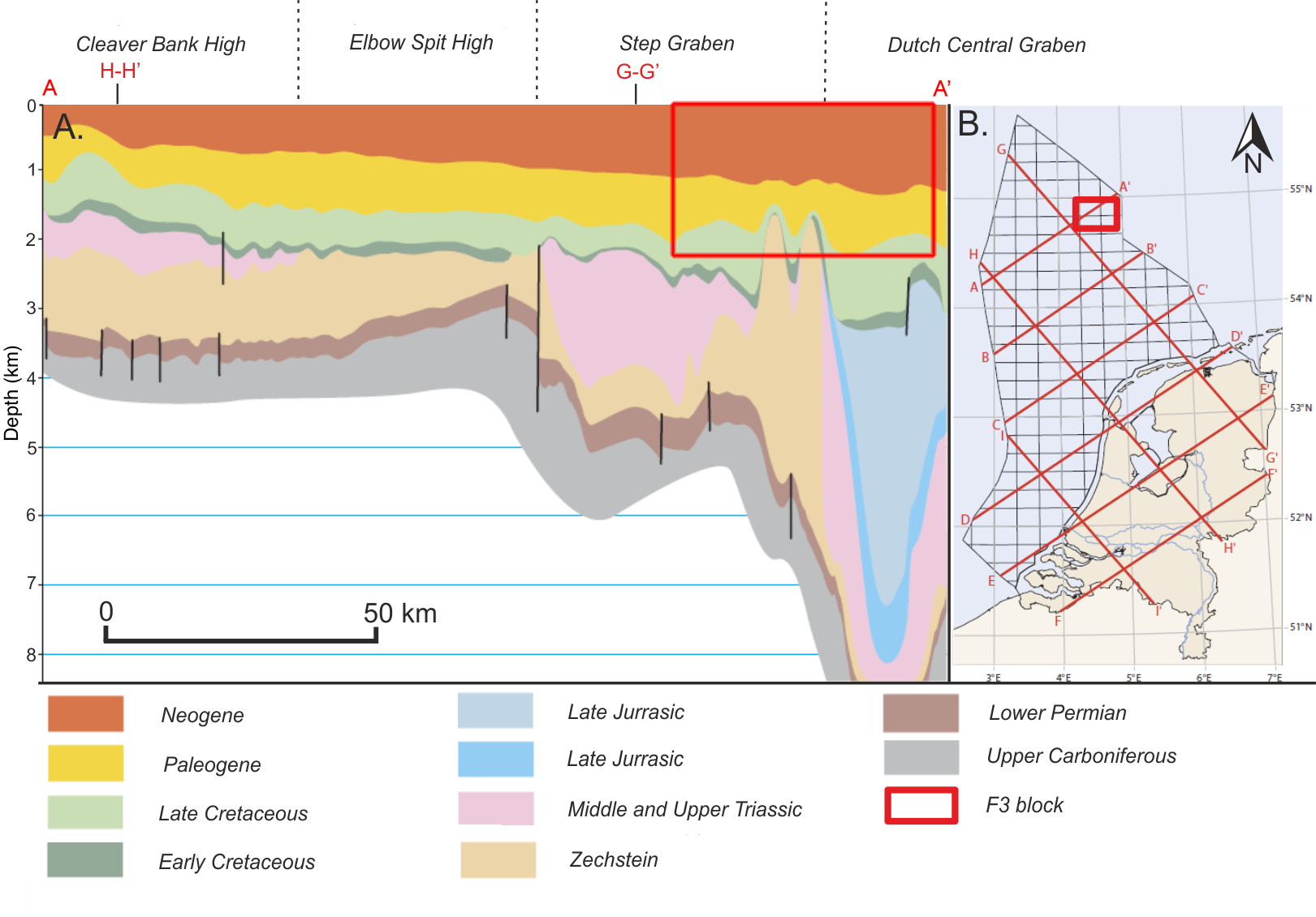}
\end{center}

\caption{A) A geological cross-section of the North Sea continental shelf along axis A-A'; B) A map of the location of the cross-section. Adapted from \cite{duin2006subsurface}.}
\label{fig:geo_cross_section}
\end{figure*}

The F3 block is located on the border of two tectonic structures: the Step Graben and the Dutch Central Graben (see Figure \ref{fig:geo_cross_section}). These tectonic structures are characterized by different lithostratigraphic units of varying thickness. This varying thickness is a result of tectonic activity \cite[]{ziegler1988evolution, ziegler1990geological}, which was started in the Variscan orogeny \cite[]{schroot2003improved}. The area within the Step Graben is strongly disturbed by salt diapirs, which were active several times, from the Zechstein to the Paleogene period \cite[]{remmelts1996salt}. On the other hand, and as a result of subsiding Jurassic rocks, the Altena, Scruff, Schieland and Niedersachsen groups are observed only within the Dutch Central Graben \cite[]{duin2006subsurface}.

\subsection{The modeling process}

To prepare our 3D geological model of the F3 block, we relied on both well logs and 3D seismic data. The next two subsections describe this process.

\subsubsection{3D model building using well logs data}

\begin{figure}[!ht]

\begin{center}
\includegraphics[width=\columnwidth]{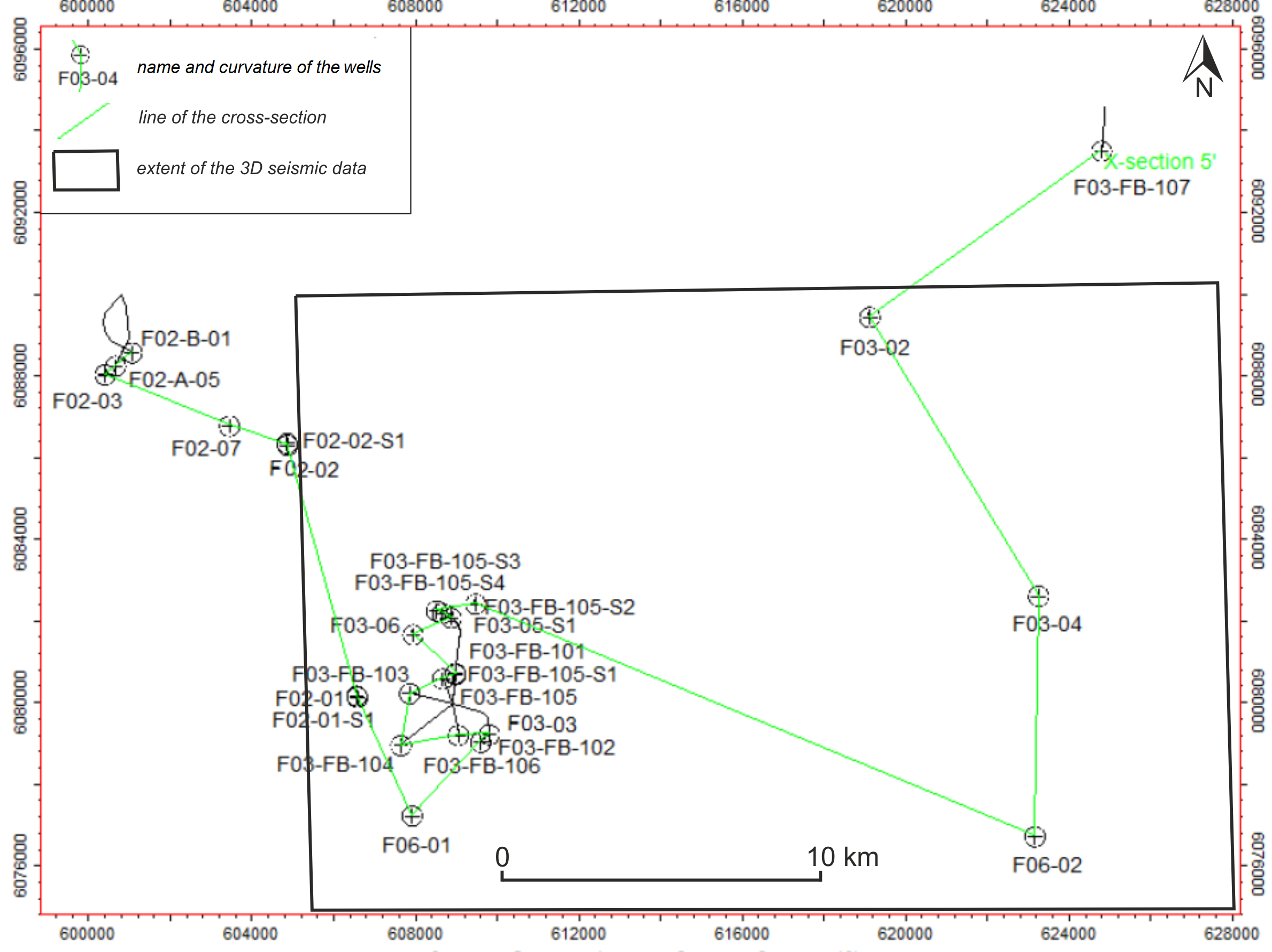}
\end{center}

\caption{Locations of the boreholes that were used to create the geological model.}
\label{fig:well_logs}
\end{figure}

The well log data were obtained from a website managed by the Geological Survey of the Netherlands (\url{www.nlog.nl}). The data (including information related to coordinates, true vertical depth, measured depth along the curvature, inclinations, and individual horizons) were collected for 26 boreholes located within the F3 block or its vicinity. The exact locations of these wells are visualized in Figure \ref{fig:well_logs}. 

Originally, the 26 wells contained 40 different horizons, so it was necessary to assign these different horizons to the various lithostratigraphic units that were adopted in literature and were presented in the previous subsection.  The next step was correlating wells with each other. After that, it was possible to create a preliminary 3D model based on the well log data by using Petrel's \textit{make/edit surface} tool. This process facilitated the preliminary visualization of the range of individual horizons, which was very helpful in the further interpretation of the 3D seismic data.

\subsubsection{3D model building using seismic data}
\begin{figure*}[ht]
\centering
\includegraphics[width=\textwidth]{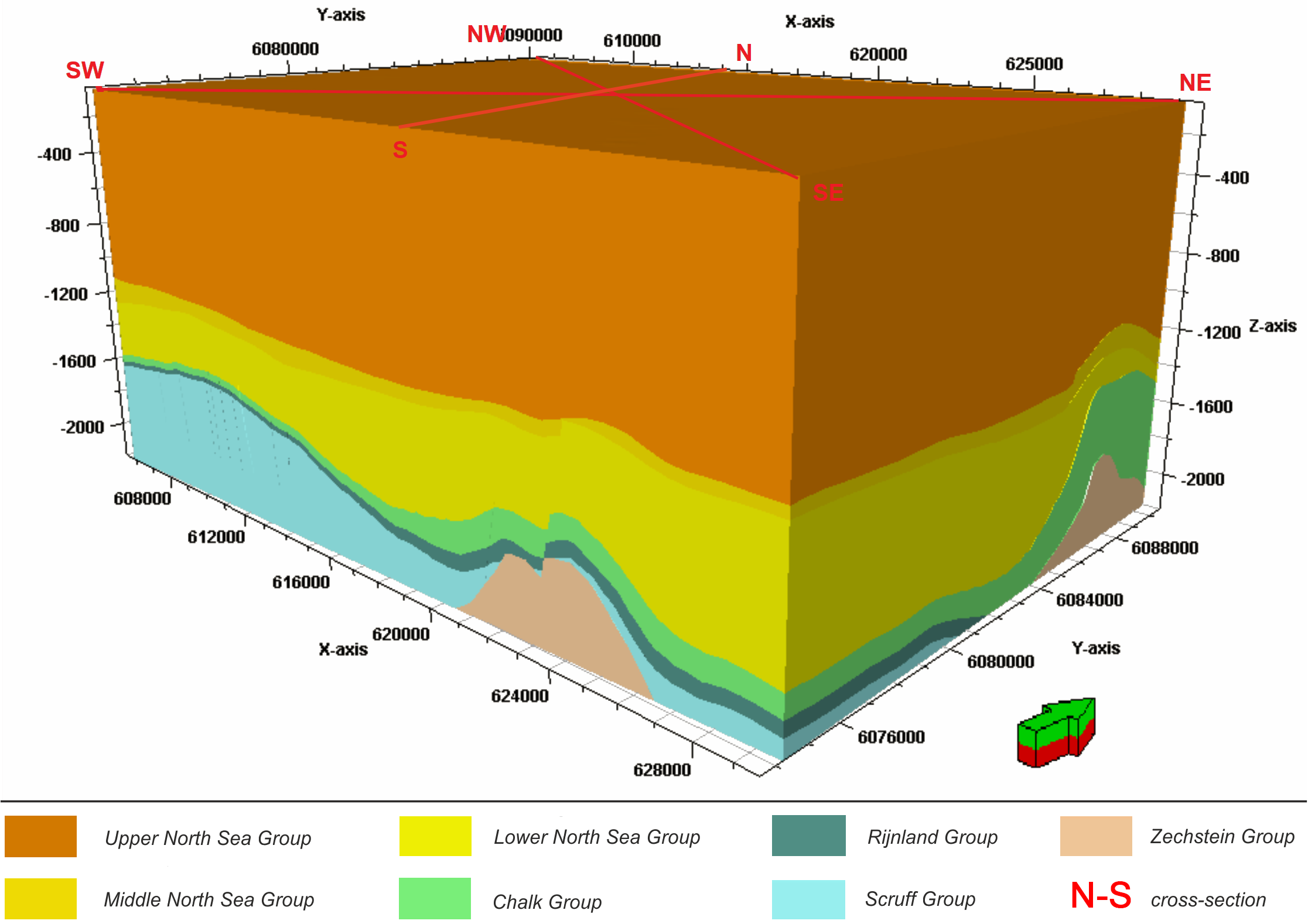}
\caption{A 3D view of our geological model of the F3 block.}
  \label{fig:model}
\end{figure*}

The F3 block data was migrated in time, not depth, so it was necessary to do time-depth conversion since the structural model must be prepared in the depth domain. OpendTect 5.0 was used to perform the time-depth conversion using a velocity model that was provided with the F3 block data. 

The next step in creating the model was faults-surface interpretation. Using Petrel's \textit{polygon editing} tool, we interpreted the main fault surfaces and the fault networks were created by using the \textit{fault framework modeling} tool. Horizons were interpreted in a similar fashion, but by using the \textit{seeded 3D autotracking} tool, which interpolated data automatically and took into account the faults networks modeled previously.

Based on the interpreted horizons and faults, preliminary modeling was conducted. This was done using the \textit{horizon modeling} tool with \textit{volume-based modeling}, which is an advanced method of isochronous geological space modeling. The preliminary model included several imperfections in the interpretations of horizons and faults, so it was necessary to re-model several faults and conduct small corrections in the interpreted horizons. 

After this, it was possible to create the final three-dimensional model which highlights the regions between individual horizons. Here, Petrel's  \textit{structural modeling} module in the \textit{horizon modeling} tool was used in addition to the \textit{create zone model} function. The final 3D geological model is shown in Figure \ref{fig:model}.

\subsection{The 3D geological model}

Within our 3D geological model of the F3 Block, we identified seven groups of lithostratigraphic units (see Figure \ref{fig:model}). These are (from newest to oldest): the Upper North Sea group, the Middle North Sea group, the Lower North Sea group, the Chalk group, the Rijnland group, the Scruff group, and the Zechstein group.

These groups can be divided into three structural levels: 
\textbf{Cenozoic} (Lower, Middle, and Upper North Sea groups), 
\textbf{Mesozoic} (Scruff, Rijnland, and Chalk groups), and 
\textbf{Permian} (Zechstein group).

As is evident in Figure \ref{fig:model}, the F3 Block is characterized by highly variable geological structures, both in the horizontal and vertical range, which is manifested by the differential thicknesses of individual units and by the expanded faults network related to salt tectonics. The area of the F3 Block can be divided into two regions: Eastern and Western. The Eastern region is disturbed by the occurrence of Zechstein diapirs and irregular faults network. The Western region is characterized by regular fault networks and a more uniform thickness of lithostratigraphic units.

The \textbf{Upper North Sea group} is the youngest and the flattest lithostratigraphic unit within our model. The top of the Upper North Sea group is the bottom of the North Sea at the same time, which is about -40 meters above sea level (m a.s.l). Differences in the depth of the ocean floor are small, and they are maximally 6 meters within the whole F3 Block. It can be noted that the depth of this top decreases from SW to NE. The thickness of the Upper North Sea group varies from about 1000 m (in places deformed by Permian diapirs) to about 1320 m in the northern part of the research area (see Figure \ref{fig:model}).

Below the Upper North Sea group lays the \textbf{Middle North Sea group}. The depth of the top of this unit ranges from -1000 m a.s.l. within the diapir in NE part of the F3 Block to about -1360 m a.s.l. in the northern part of this area, between diapirs. The thickness of the Middle North Sea group is from 20 to 150 m. As in the case of the Upper North Sea group, there is a clear relationship between the occurrence of Zechstein salts and the depth and thickness of this unit. Differences in the thickness of this unit between both sites of faults are also visible. 

The next unit is the \textbf{Lower North Sea group}. This unit contains similar lithostratigraphic units to the Middle North Sea group, but is visually distinct in the seismic data. The top is at a depth from -1100 m a.s.l., while the thickness is from about 180 to 750 m. 

The top of the \textbf{Chalk group} is at depth from -1300 m a.s.l. (above the diapirs in the NE part of the survey) to -2100 m a.s.l. (in the Eastern part of the survey, which is undisturbed by diapirs). The minimum thickness of this unit is 25 m, while above the salt diapirs in NE part of the F3 Block, this substantially increases to 525 m. 

\begin{figure}[!ht]

\begin{center}
\includegraphics[width=\columnwidth]{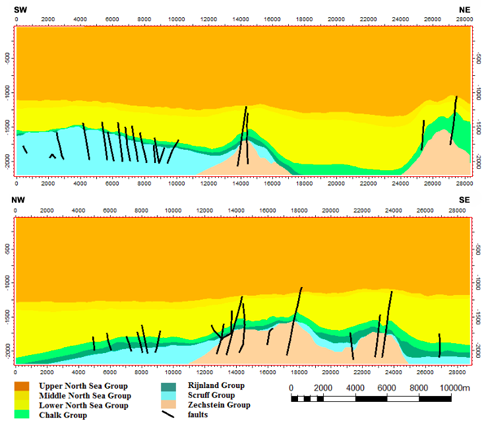}
\end{center}

\caption{Two diagonal cross sections of our 3D geological model in Figure \ref{fig:model}.}
\label{fig:model_cross}
\end{figure}

The \textbf{Rijnland group} is submerged in the NNE direction, while it is the shallowest in the SW part of the F3 Block and above some Zechstein diapirs at the center of the survey (see Figure \ref{fig:model_cross}). The maximum thickness of the Rijnland group is about 200 m (above some diapirs), while in the other parts of the F3 block it can be less than 20 m or does not occur at all.

The \textbf{Scruff group}, similar to the Rijnland group, is thinned out in NNE direction, more or less in the middle of the F3 Block, where the top of this layer has a depth of -2180 m a.s.l. This layer is shallowest (-1500 m a.s.l.) in the SW part of the F3 block and above the Zechstein diapirs in the Southern part of the survey. The thickness of the Scruff group within our model boundaries ranges from 100 m to almost 700 m, but is much larger in reality and can reach several kilometers \cite[]{duin2006subsurface}.

The \textbf{Zechstein group} occurs only in the eastern part of the survey, as irregularly-shaped salt diapirs. The shallowest part of the Zechstein group is at a depth of -1500 m a.s.l. while the maximum thickness of the Zechstein group within the research area is about 700 m. However, as in the case of the Scruff group, the depth is much bigger. According to the literature, it can reach several kilometers \cite[]{duin2006subsurface}.

\begin{figure}[!ht]

\begin{center}
\includegraphics[width=\columnwidth]{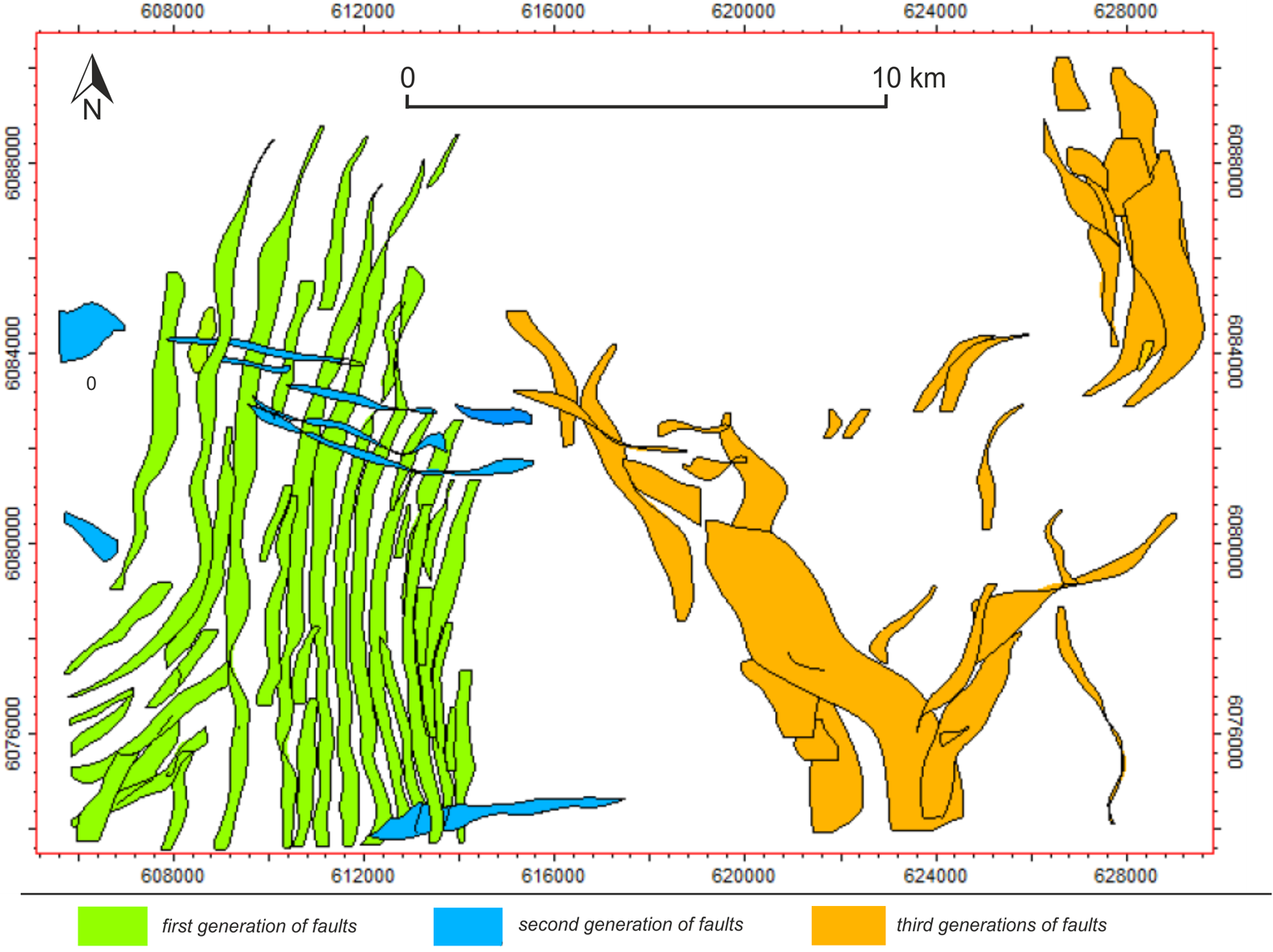}
\end{center}

\caption{An overhead view of 3D fault planes from three different generations of faults that we have identified in the F3 block.}
\label{fig:faults}
\end{figure}

In addition to the identified groups of lithostratigraphic units mentioned above, we have also identified three generations of \textbf{faults}. The first generation are reverse, oblique-slip, sinistral faults with an SSW-NNE orientation. This direction is connected with the course of the tectonic axis of the Dutch Central Graben, which (similar to the whole Graben) has an SSW-NNE orientation. The second generation of faults are normal, oblique-slip, dextral faults with a W-E orientation. Finally, the third generation are faults that are genetically linked with faults from the first and second generations, but were disturbed by the Permian halokinesis. Figure \ref{fig:faults} shows an overhead view of the three generations of faults that we have identified. Also, Figure \ref{fig:model_cross} shows two diagonal cross sections along the SW-NE and NW-SE axis in our 3D model shown in Figure \ref{fig:model}.

\section{Deconvolution Network Baseline}

In addition to the geological model we have introduced in the last section, we propose two baseline models for facies classification based on a deconvolution network architecture. In this section, we introduce deconvolution networks and describe our two baseline models. 

\subsection{Deconvolution networks}
\begin{figure*}[ht!]

\begin{center}
\includegraphics[width=0.8\textwidth]{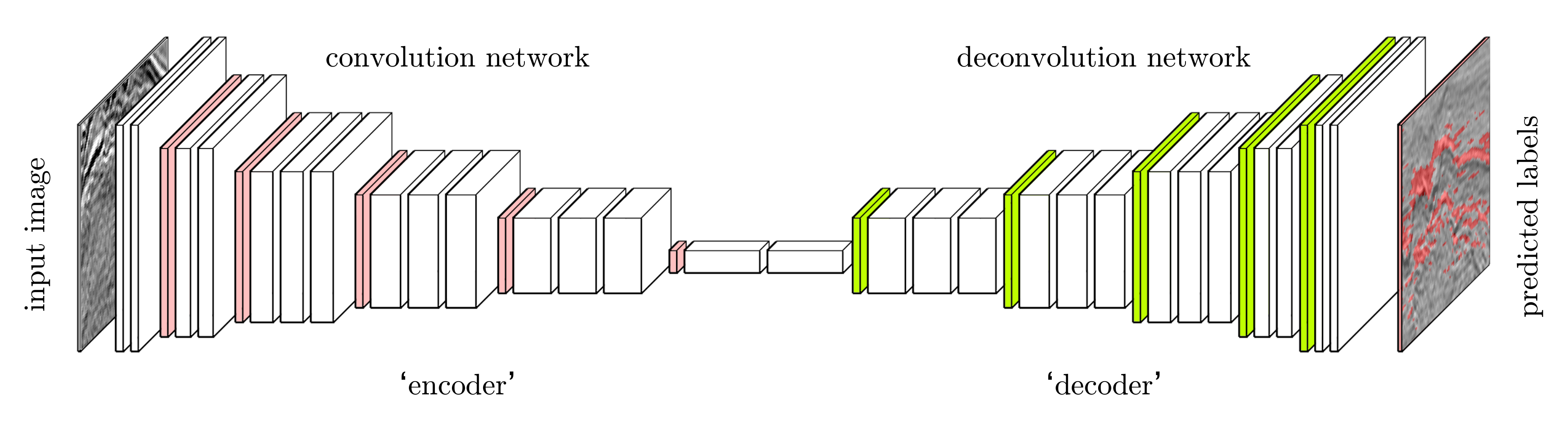}

\end{center}

\caption{The architecture of the deconvolution network used in this work. White layers are convolution or deconvolution layers. Red layers are max-pooling layers, while green layers are unpooling layers.}
\label{fig:deconv}
\end{figure*}

Convolutional neural networks (CNNs) have seen great success in a wide number of visual applications, from image classification to semantic labeling. In the previous few years, it was not well established that end-to-end CNN architectures can perform very well in semantic labeling tasks (such as facies classification). A major hurdle to the success of end-to-end CNN architectures in these tasks was what seemed like a trade-off between classification and localization accuracy. Deeper networks that have many convolution and pooling layers have proven to be the most successful models in image classification tasks. However, their large receptive fields and increased spatial invariance (due to pooling and convolutional layers) make it difficult to infer the locations of various objects within the image. In other words, the deeper we go into a network, the more it seemed we lose the location information of various objects within the image. Some researchers have attempted to overcome this hurdle by using various pre- or post-processing techniques. However, the introduction of fully convolutional network architectures, such as FCN \cite[]{long2015fully} and DeconvNet \cite[]{noh2015learning} have shown that it is possible to achieve good semantic labeling results using a convolutional network only, with no pre- or post-processing steps required. FCN accomplish this by replacing the fully-connected layers of the CNN with 1D convolutional layers that produce coarse feature maps. These coarse feature maps are then upsampled, and concatenated with the scores from intermediate feature maps in the network to generate the output. These upsampling steps, however, result in a blurred output that loses some of the resolution of the original image. 

Deconvolution networks, on the other hand, overcome this problem by using a symmetric encoder-decoder style architecture composed of stacks of convolution and pooling layers in the encoder, and stacks of deconvolution and unpooling layers in the decoder that mirror the encoders architecture. The role of the encoder can be seen as doing object detection and classification, while the decoder is used for accurate localization of these objects within the image. This architecture can achieve finer and more accurate results than those of the FCN, and therefore is adopted in our work.

A few recent papers have illustrated the 
successful application of deconvolution networks for seismic interpretation applications \cite[]{yazeed_seg2018_label, haibin_deconv_seg2018}.

Figure \ref{fig:deconv} illustrates the architecture of the deconvolution network used for both of our baseline models. Every convolution or deconvolution layer (in white) is followed by a rectified linear unit (ReLU) non-linearity. The layers in red perform $2\times2$ max pooling to select the maximum filter response within small windows. The indices of the maximum responses for every pooling layer are then shared with their respective unpooling layers (in green) to undo this pooling operation and get a higher resolution image. 

\subsection{Baseline Models}
In this work, we use two baseline models; a patch-based and a section-based model. These two models use the exact same architecture, optimizer, and hyperparameters but differ in the way they are trained and the way they are used to label the seismic volume. 

\subsubsection{Patch-based model:}

The patch-based model is trained on small patches extracted from the inlines and crosslines of the training data. For very large seismic volumes, this approach can be more feasible than using entire sections for training. At training time, the patches of seismic data and their associated labels are sampled randomly from the inlines and crosslines of the training set. During test time, the model samples overlapping patches in the inline and crossline direction and averages the results to generate a 2D labeled version of the test inline or crossline. This is done for all inlines and crosslines in the test sets. 

\subsubsection{Section-based model:}
The section-based model is trained on entire inline and crossline sections. The advantage of this approach is two-fold. First, since the network is fed an entire section, it can easily learn the relationships between different lithostratigraphic units and can take the depth information into account when labeling the section. The second advantage is more practical. Training and testing entire sections at once means the network can be trained or tested very quickly since there are only a relatively small number of seismic inlines and crosslines\footnote{This is assuming the GPU memory is large enough to handle the size of the seismic sections. On our \textsc{Nvidia} Titan X GPU, we trained the baseline section-based network -- eight sections at a time -- in about 70 minutes.}. One advantage of using a fully convolutional architecture (such as the one we are using) is that the size of the network input does not have to be fixed. The size of the output of the network changes as the size of its input change. Therefore, the different size of the inline and crossline sections does not pose any problem to the training of this network\footnote{While the sizes of the inlines and crosslines do not need to match, their resolutions (in terms of meters/pixel) should. In our case, pixels in the inline and crossline directions are both 25m$\times$25m.}.

\subsubsection{Other variations:}
In addition to the baseline patch- and section-based models, we have trained other variations of these models to test how they can be improved. We have tested the following variations: 

\begin{itemize}
    \item \textit{Baseline + data augmentation:} data augmentation applies different label-preserving transformations to the training data such as small rotations, random horizontal flipping, and the addition of Gaussian noise. This can help increase the training sample size, and help the network generalize better to the test data. 
    \item \textit{Baseline + data augmentation + skip connections:} we further improve on the previous model by adding skip connections. In a deep neural network, the output of a layer is typically passed on as the input to the next layer in the network. Skip connections allow the output of a layer to be also passed as an input to a layer farther up the network, skipping intermediate layers in the process. These connections are implemented by directly adding the outputs of various layers in the encoder part of the deconvolution network to the outputs of the corresponding layers in the decoder. Skip connections help networks overcome the vanishing gradient problem \cite[]{hochreiter2001gradient} by providing ``shortcuts'' for the computed gradients to propagate to the lower layers of the network. 
    
\end{itemize}

\section{Experimental Setup}
\begin{figure*}[ht!]
\centering
\includegraphics[width=\textwidth]{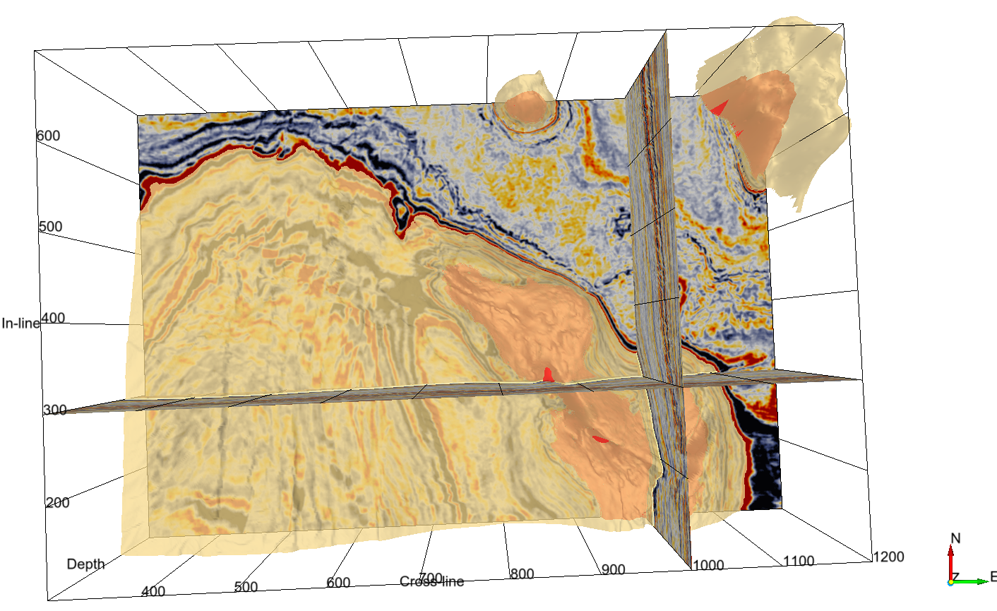}
\caption{A 3D view of the F3 block from above with the \hlc[red]{  Zechstein  } Group shown in red, while the \hlc[rijnchalk]{  Rijnland/Chalk  }  Group is shown in a semi-transparent beige color. Inline 300 and crossline 1000 divide the survey into four regions. The NW region of the survey is used for training, while the SW region constitutes test set \#1. The remaining region East of crossline 1000 constitutes test set \#2.}
  \label{fig:survey_split}
\end{figure*}

In this section, we will introduce the main elements of the experimental setup, including how the final geological model was produced, how the model is split into training and testing sets, and what metrics are used to objectively evaluate the performance. 

\subsection{The geological model}

The final geological model that we use to train and test our models is not the entire volume shown in Figure \ref{fig:model}. The time-depth conversion process of the seismic data resulted in some artifacts. These artifacts were concentrated along the sigmoidal structure in the Upper North Sea group. Due to these artifacts, and missing data on the sides of the survey, we only use the data between inlines 100 and 701, crosslines 300 and 1201, and depth between 1005 and 1877 meters. Furthermore, we combine the Rijnland and the Chalk groups in our final model to a single class due to various issues with processing the Rijnland/Chalk boundary when generating the final model. Table \ref{tab:percent} shows the percentage of different classes in our training set. 

In addition to the final model labels and seismic data, we also release the original horizons for all the lithostratigraphic units, in addition to the extracted fault planes from all three generations.

\begin{table*}[]
\centering
\caption{The percentage of pixels from different classes in the training set.}
\def\arraystretch{1.2}
\label{tab:percent}
\begin{tabular}{|cccccc|}
\hline 
\multicolumn{1}{c}{Zechstein} & \multicolumn{1}{c}{Scruff} & \multicolumn{1}{c}{Rijnland/Chalk} & \multicolumn{1}{c}{Lower N. S.} & \multicolumn{1}{c}{Middle N. S.} & \multicolumn{1}{c}{Upper N. S.} \\ \hline 
\multicolumn{1}{c}{1.48\%}      & \multicolumn{1}{c}{3.17\%}  & \multicolumn{1}{c}{6.53\%}          & \multicolumn{1}{c}{48.44\%}      & \multicolumn{1}{c}{11.89\%}       & \multicolumn{1}{c}{28.49\%}  \\ 
\hline 
\end{tabular}
\end{table*}

\subsection{The train/test split}

Careful selection of the training and testing sets is crucial in any machine learning application. This is especially true in seismic data, where neighboring sections are highly correlated. Selecting the training and testing sections randomly will lead to artificially good test results, that are not representative of the actual generalization performance of the tested models. Therefore, it is important to minimize the correlation between the training and testing sets as much as possible. It is also important to ensure that both the training and testing sets have adequate representation of all the classes in the dataset. 

Therefore, we decide to split the data as shown in Figure \ref{fig:survey_split}. Namely, the data is split into the following three sets: 

\begin{enumerate}
    \item \textbf{Training set:} This includes all the data in the range of inlines [300,700] and crosslines [300,1000].
    \item \textbf{Test set \#1:}  This set includes all the data in the range of inlines [100,299]  and crosslines [300,1000].
    \item \textbf{Test set \#2:} This sets includes all the data in the ranges of inlines [100,700] and crosslines [1001,1200]. This set includes a large Zechstein diapir in the NE of the survey that is never seen in the training set.  
\end{enumerate}

For a fair comparison with others who might use this benchmark in the future, it is important to note that the test sets \textit{should not be used more than once}. Testing a model on the test set, then retraining that model with different parameters means that the test set has been used for validation, which defeats its purpose. 

\subsection{Evaluation metrics}

To objectively evaluate the performance of different models on our two test sets, we use the following metrics: pixel accuracy (PA), class accuracy (CA) for each individual class, mean class accuracy (MCA) for all classes, and frequency-weighted intersection over union (FWIU). These metrics are detailed in the appendix.

\section{Results}
\begin{table*}[ht]
\centering
\caption{Results of our two baseline models, with other variations, when tested on both test splits of our dataset. All metrics are in the range $[0,1]$, with larger values being better. The best performing model for every metric is highlighted in bold.}
\label{tab:results}
\resizebox{\textwidth}{!}{%
\bgroup
\def\arraystretch{1.4}
\begin{tabular}{|l||c|cccccc|c|c|}
    \hline
     \multirow{2}{*}{\diagbox[innerwidth=4.5cm]{\textbf{Model}}{\textbf{Metric}}}   & \multirow{2}{*}{\textbf{PA}} & \multicolumn{6}{c|}{\textbf{Class Accuracy}} & \multirow{2}{*}{\textbf{MCA}} & \multirow{2}{*}{\textbf{FWIU}}  \\ 
    & & \cellcolor{zechstein}Zechstein & \cellcolor{scruff}Scruff  & \cellcolor{rijnchalk}Rijnland/Chalk & \cellcolor{lowerns}Lower N. S. & \cellcolor{middlens}Middle N. S. & \cellcolor{upperns}Upper N. S. &  &  \\ \hline
    Patch-based model  & 0.788 & 0.264 & 0.074 &0.499  & 0.992 &0.804 & 0.754&0.565 &  0.640  \\ 
    Patch-based + aug. &0.852  &  0.434 & 0.221  & 0.707 & 0.974 & 0.884 & 0.916 &0.689 & 0.743\\ 
    Patch-based + aug + skip & 0.862 &0.458  &0.286  & 0.673 &0.974  &0.912&0.926 & 0.705& 0.757\\
    \hline
    \hline 
    Section-based model   & 0.879  & 0.219  &  0.539  & 0.744  & 0.951  &  0.872 &   0.973 & 0.716  & 0.789 \\ 
    Section-based + aug. &  0.901 & 0.714 &  0.423 &  0.812&  0.979 &  0.940 &0.956 & 0.804 &0.844  \\  
    Section-based + aug + skip & 0.905 & 0.602  &   0.674 & 0.772 & 0.941 & 0.938 & 0.974 & 0.817 & 0.832 \\
    
    \hline
    \end{tabular}
    \egroup
}
\end{table*}

\begin{figure*}[!htb]
  \centering

    \subfloat[Seismic data]{\includegraphics[width=0.49\textwidth]{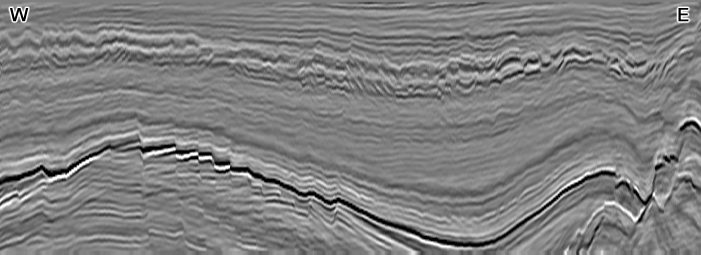}}
    \hspace{5pt}
    \subfloat[Ground truth labels]{\includegraphics[width=0.49\textwidth]{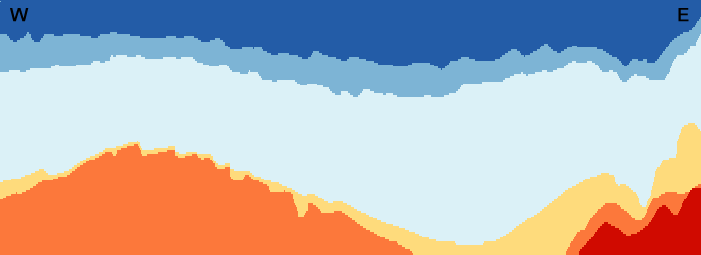}}

    \subfloat[Patch-based baseline]{\includegraphics[width=0.49\textwidth]{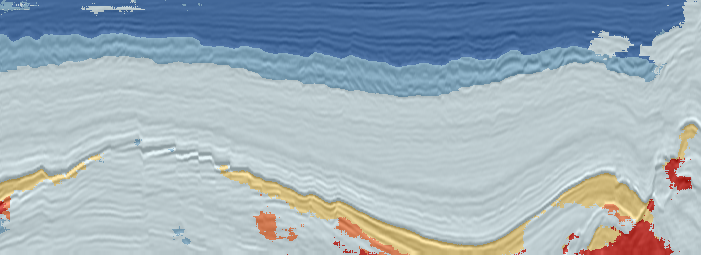}}
    \hspace{5pt}
    \subfloat[Section-based baseline]{\includegraphics[width=0.49\textwidth]{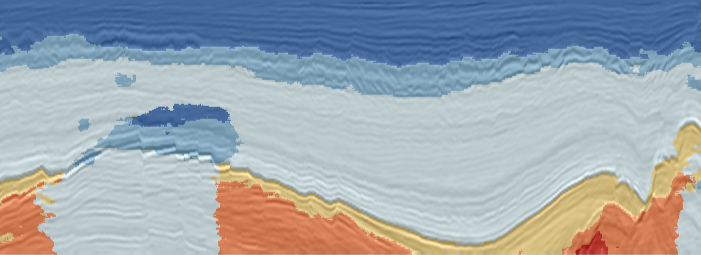}}
    
    \subfloat[Patch-based + aug]{\includegraphics[width=0.49\textwidth]{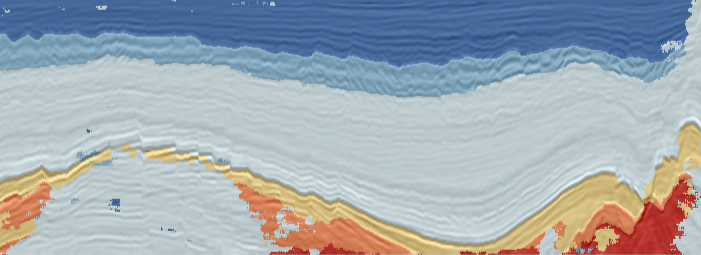}}
    \hspace{5pt}
    \subfloat[Section-based + aug]{\includegraphics[width=0.49\textwidth]{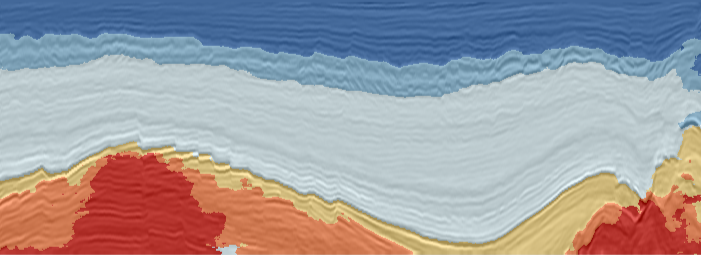}}
    
    \subfloat[Patch-based + aug + skip]{\includegraphics[width=0.49\textwidth]{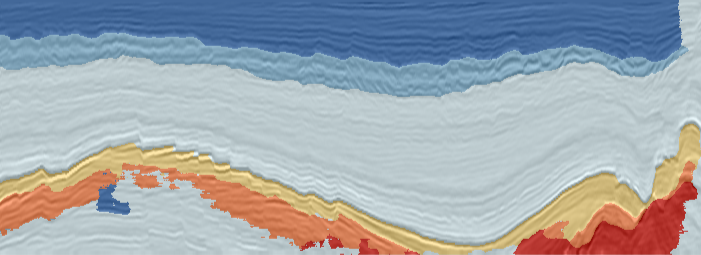}}
    \hspace{5pt}
    \subfloat[Section-based + aug + skip]{\includegraphics[width=0.49\textwidth]{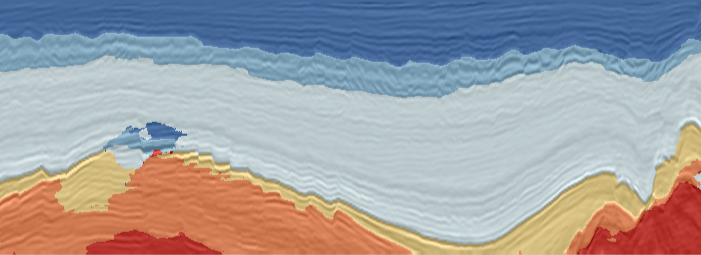}}

  \caption{The results of the different models on inline 200 from test set \#1. The color map is shown in Table \ref{tab:results}.}
  \label{fig:results}
\end{figure*}

After the final geological model was created, we train each of the models described earlier on the training set until its training loss converges. We note that on our \textsc{Nvidia} Titan X GPU, the baseline patch-based model, and the augmented version, converged after 16 hours of training. The patch-based model with skip connections required less than 5 hours to converge. The section-based models required significantly less time, all of them converging in less than 90 minutes. We test these models by using them to label all inlines and crosslines in both test sets, and computing the performance metrics on the final result. Table \ref{tab:results} summarizes the objective results for all the models that we have tested on both test sets\footnote{The results for test set \#1 and test set \#2 are available from \url{https://bit.ly/2D4QTbH}}, while Figure \ref{fig:results} shows inline 200 of test set \#1 labeled using the six different models we have tested. In the remainder of this section, we will discuss these results, and suggest various methods to improve upon them. 

\subsubsection{Patch-based vs section-based models:}

\begin{figure}[!t]
\centering
\subfloat[Patch-based baseline model]{\includegraphics[width=0.5\textwidth]{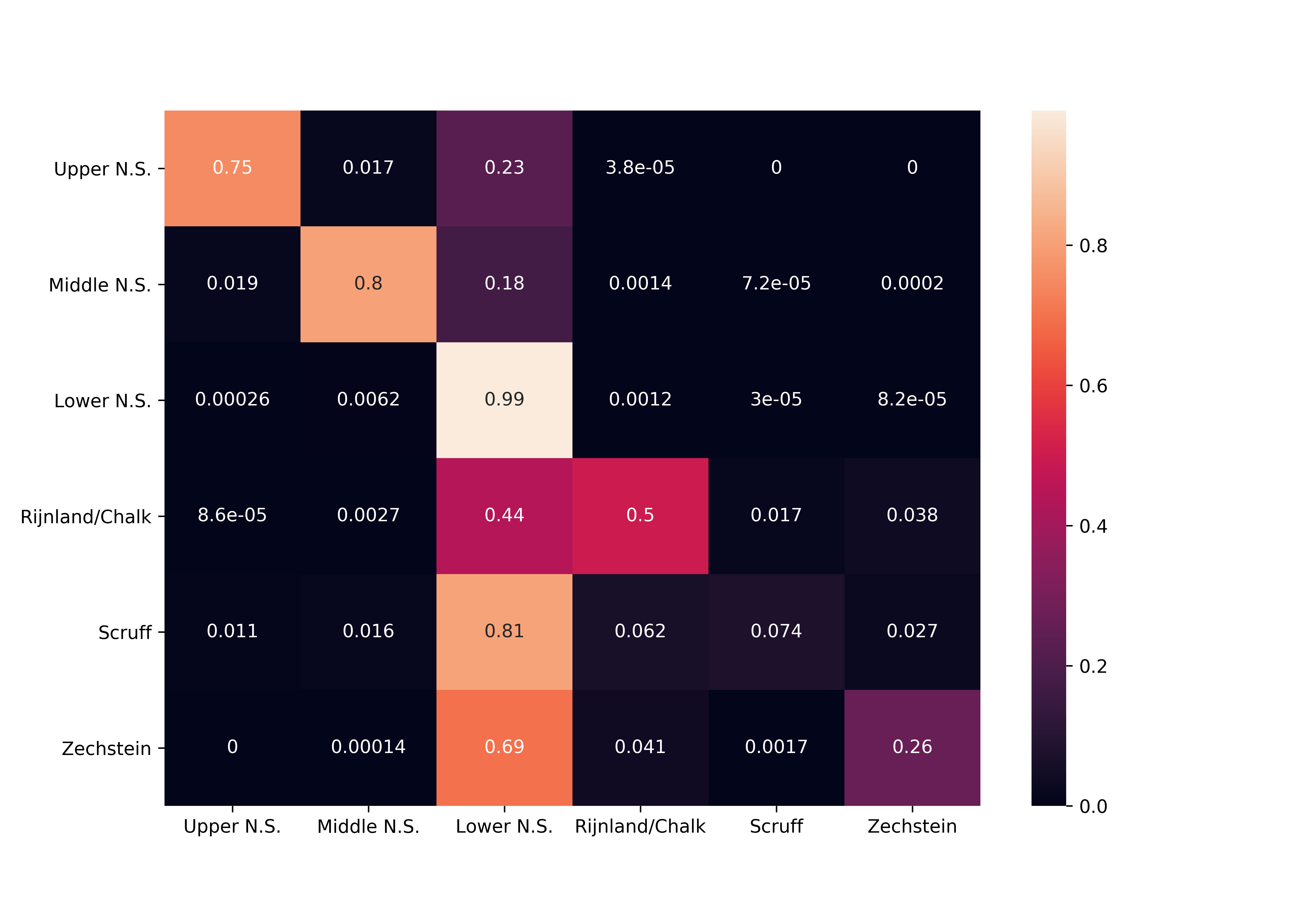}}
\\
\subfloat[Section-based baseline model]{\includegraphics[width=0.5\textwidth]{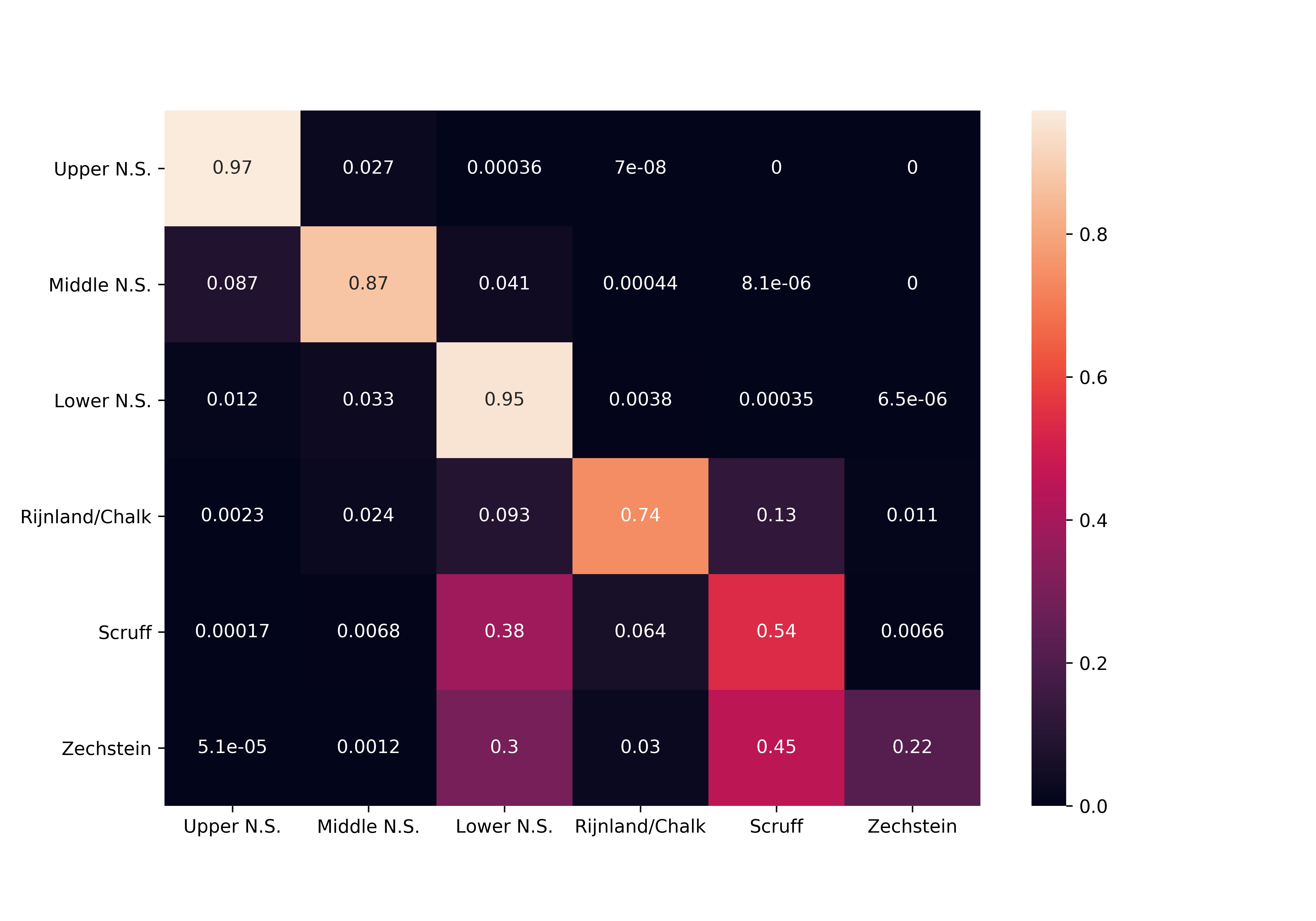}}
\caption{Confusion matrices for our two \textit{baseline} models on both test set \#1 and \#2. Each row shows the distribution of the model output for each class.}
\label{fig:confusion}
\end{figure}

Since the patch-based models are trained on patches from different depths in the data, they can easily confuse various classes that typically exist at different depths. For example, the patch-based models in Figure \ref{fig:results} often confuse the Scruff group in the bottom left of the image with the Lower North Sea group, while the section-based models do not suffer from these problems as often. Figure \ref{fig:confusion} shows the confusion matrices for the baseline patch and section models. It shows how the patch-based model confuses many classes in our test sets with the Lower North Sea group. The baseline section-based model is better at classifying the other classes as Figure \ref{fig:confusion}(b) shows. 

Table \ref{tab:results} shows that both patch and section-based models perform fairly well on the North Sea groups, with the section-based models performing better. However, for smaller classes such as the Scruff and Zechstein groups, the section-based models show a clear advantage. The MCA score shows a 15\% improvement of the section-based baseline model vs. the patch-based model. Overall, section-based models are superior to patch-based models due to their ability to incorporate spatial and contextual information within each seismic section. They also have the advantage of being faster to train and test. 

\subsubsection{Imbalanced classes:}
As Table \ref{tab:percent} shows, our dataset is highly imbalanced. The Zechstein and Scruff groups are far smaller than the Lower or Upper North Sea groups. This means while training, the network is trained on far more examples of Lower or Upper North Sea groups than of Zechstein or Scruff groups for example. This leads to the networks being biased towards classifying pixels as Lower or Upper North Sea groups, and therefore artificially achieve high class accuracy scores for those classes. However, this is at the expense of the very poor performance for the smaller classes. In addition, the larger classes have a more diverse visual appearance compared to smaller ones, and therefore, it would be easier for the network to confuse features from smaller classes with those learned from larger ones. 

We do not make any changes to the baseline models to overcome this imbalance. However, using various techniques to overcome this class imbalance can significantly improve the results, especially for the smaller classes, such as Zechstein and Scruff groups.

\subsubsection{Data augmentation:}
Data augmentation is a technique to increase the size of the training set artificially. This is quite useful when training a large network with a limited amount of training data. We use simple augmentation operations including randomly rotating the patch or the section by up to $\pm 15^\circ$, adding random Gaussian noise, and randomly flipping the patches or the sections horizontally. Using data augmentation significantly improved the results for both baseline models, but especially the patch-based model. The FWIU and MCA scores increased by more than 10\% in the patch-based model, and significantly improved the results for smaller classes such as the Zechstein and Scruff groups. The results of the section-based model were also improved by using data augmentation, although to a lesser degree. 

\subsubsection{Skip connections:}
For both patch- and section-based models, adding skip connections can improve the results, and speed up the training. This is especially noticeable in the patch-based model where adding skip connections to the \textit{baseline + aug} model improved the results by about 1\% in the PA metric and about 1.5\% in the MCA and FWIU metrics. The improvement in the results in the section-based models are more subtle, as adding skip connections only improved the PA result by 0.1\%. Interestingly, the Scruff group which worst is the performing class in both the patch and section-based model seemed to benefit the most from the addition of skip connections. The class accuracy score for the Scruff group increased by 6.5\% and 25\% in the patch and section based models respectively. Overall, adding skip connections seems to help improve the results. It also can speed up to the training process. In the case of the patch-based model, the skip connection model converged four times faster than the baseline.

\section{Conclusions}

In conclusion, we have introduced and made publicly available a new annotated dataset for facies classification. This dataset includes six different lithostratigraphic classes based on the underlying geology of the Netherlands F3 block. The dataset also includes fault planes from three different generations that we have identified in the F3 block. In addition, we present two baseline deep learning models for facies classification, a patch- and a section-based model, both based on a deconvolution network architecture. We train these models using our dataset, and we analyze their performance. Furthermore, we make the code for training and testing these models publicly available for others to use. It is our hope that this dataset, and the code that we have released, will help facilitate more research in this area, and help create an objective benchmark for comparing the results of different machine learning approaches for facies classification. 

\section{Acknowledgements}
We want to acknowledge the support of the Center for Energy and Geo Processing (CeGP) at the Georgia Institute of Technology and King Fahd University of Petroleum and Minerals (KFUPM). We would also like to thank dGB Earth Sciences for making the Netherlands F3 data publicly available, and Schlumberger for providing educational licenses for the Petrel software.

\append{Evaluation Metrics}
\label{appendix:eval_metrics}
To objectively evaluate the performance of our models on this dataset, we use a set of evaluation metrics that are commonly used in the computer vision literature. If we denote the set of pixels that belong to class $i$ as $G_i$, and the set of pixels classified as class $i$ as $F_i$. Then, the set of correctly classified pixels is $G_i \cap F_i$. We use $|\cdot|$ to denote the number of elements in a set. Now, we can define the following metrics:

\begin{itemize}
    \item \textbf{Pixel Accuracy (PA)} is the percentage of pixels over all classes that are correctly classified,
    \begin{equation}
    \text{PA} = \frac{\sum_i |F_i\cap G_i|}{\sum_i |G_i|}.
    \end{equation}
    
    \item \textbf{Class Accuracy for class $i$  ($\text{CA}_i$)} is the percentage of pixels that are correctly classified in a class $i$. 
    \begin{equation}
     \text{CA}_i = \frac{|F_i\cap G_i|}{|G_i|}.
     \end{equation}
     We will also define the \textbf{Mean Class Accuracy (MCA)} as the average of CA over all classes, 
    \begin{equation}
    \text{MCA} = \frac{1}{n_c}\sum_i \text{CA}_i =\frac{1}{n_c}\sum_i \frac{|F_i\cap G_i|}{|G_i|},
    \end{equation} 
    where $n_c$ is the number of classes.
    
    \item \textbf{Intersection over Union ($\text{IU}_i$)} is defined as the number of elements of the intersection of $G_i$ and $F_i$ over the number of elements of their union set, 
    \begin{equation}
    \text{IU}_i =\frac{|F_i\cap G_i|}{|F_i\cup G_i|}.
    \end{equation}
    This metric measures the overlap between the two sets and it should be $1$ if and only if all pixels were correctly classified. Further, when we average IU over all classes, we arrive at the Mean Intersection over Union (Mean IU),
    \begin{equation}
    \text{Mean IU} =\frac{1}{n_c}\sum_i \text{IU}_i =\frac{1}{n_c}\sum_i \frac{|F_i\cap G_i|}{|F_i\cup G_i|}.
    \end{equation}

    To prevent this metric from being overly sensitive to small classes, it is common to weigh each class by its size. The resulting metric is known as \textbf{Frequency-Weighted Intersection over Union (FWIU)},
    \begin{equation}
    \text{FWIU} = \frac{1}{\sum_i |\mathcal{G}_i|}\cdot \sum_i |\mathcal{G}_i|\cdot\frac{|\mathcal{F}_i\cup \mathcal{G}_i|}{|\mathcal{F}_i\cap \mathcal{G}_i|}.
    \end{equation}

\end{itemize}

\bibliographystyle{seg}  
\bibliography{main.bib}

\end{document}